\documentclass[epjCONF]{svjour}
\usepackage{graphics}
\usepackage[varg]{txfonts} 
\usepackage{hyperref}
\usepackage[latin1]{inputenc}
\newcommand{\diamonds}{\textsc{D\large{iamonds}}}
\session-title{Conference Title, to be filled}
\begin{document}
\title{\textsc{D{\Large{iamonds}}}: A new Bayesian nested sampling tool\thanks {Software package available at the \diamonds\,\,code website: \href{https://fys.kuleuven.be/ster/Software/Diamonds/}{https://fys.kuleuven.be/ster/Software/Diamonds/}.}}
\author{Enrico Corsaro\inst{1}\fnmsep\thanks{\email{emncorsaro@gmail.com}} \and Joris De Ridder\inst{1}}
\institute{Instituut voor Sterrenkunde, KU Leuven, Celestijnenlaan 200D, B-3001 Leuven, Belgium}
\abstract{
In the context of high-quality asteroseismic data provided by the NASA Kepler Mission, we developed a new
code, termed \diamonds\,\,(high-DImensional And multi-MOdal NesteD Sampling), for fast Bayesian
parameter estimation and model comparison by means of the Nested Sampling Monte Carlo (NSMC) algorithm,
an efficient and powerful method very suitable for high-dimensional problems (like the peak bagging analysis
of solar-like oscillations) and multi-modal problems (i.e. problems that show multiple solutions).
We applied the code to the peak bagging analysis of solar-like oscillations observed in a challenging F-type
star. By means of \diamonds\,\,one is able to detect the different backgrounds in the power spectrum of the star
(e.g. stellar granulation and faculae activity) and to understand whether one or two oscillation peaks can be
identified or not. In addition, we demonstrate a novel approach to peak bagging based on multi-modality,
which is able to reduce significantly the number of free parameters involved in the peak bagging model. This
novel approach is therefore of great interest for possible future automatization of the entire analysis technique.
} 
\maketitle
\section{The \textsc{D{\Large{iamonds}}} code}
\label{intro}
\diamonds\,\,is a software package developed in C++11 and structured in classes in order to be
as much flexible and configurable as possible \cite{Corsaro14}. It implements a more
recent version of the NSMC algorithm \cite{Skilling04,Shaw07,Feroz08,Feroz09}. The user can supply its own
likelihoods, priors and models, according to the astrophysical problem of
interest, by using a starting template. All the free parameters of the model and its
corresponding Bayesian Evidence are therefore estimated by the code.

\section{Peak Bagging and Bayesian model comparison}
\label{sec:1}
\begin{figure}
\centering
\resizebox{0.95\columnwidth}{!}{%
 \includegraphics{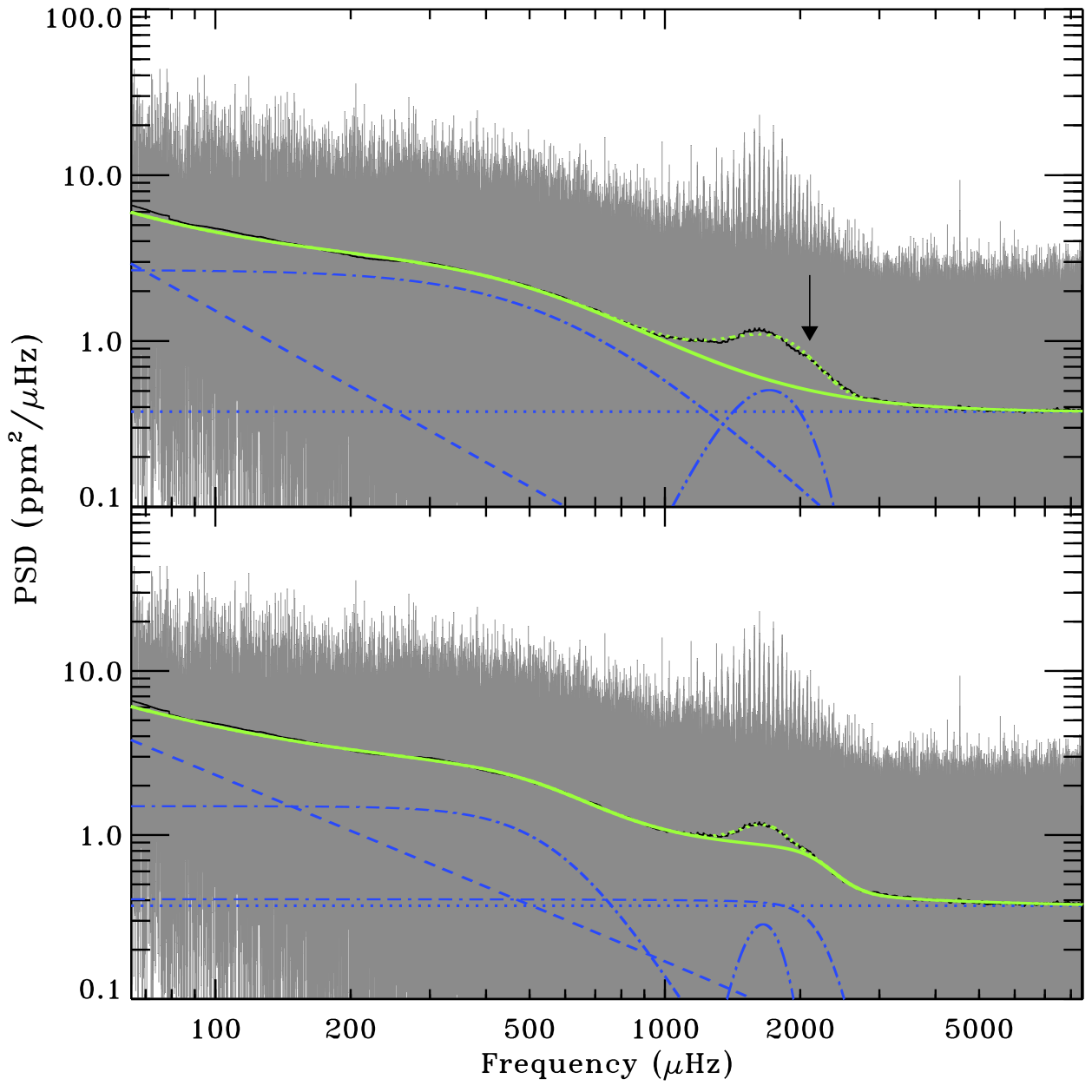}\includegraphics{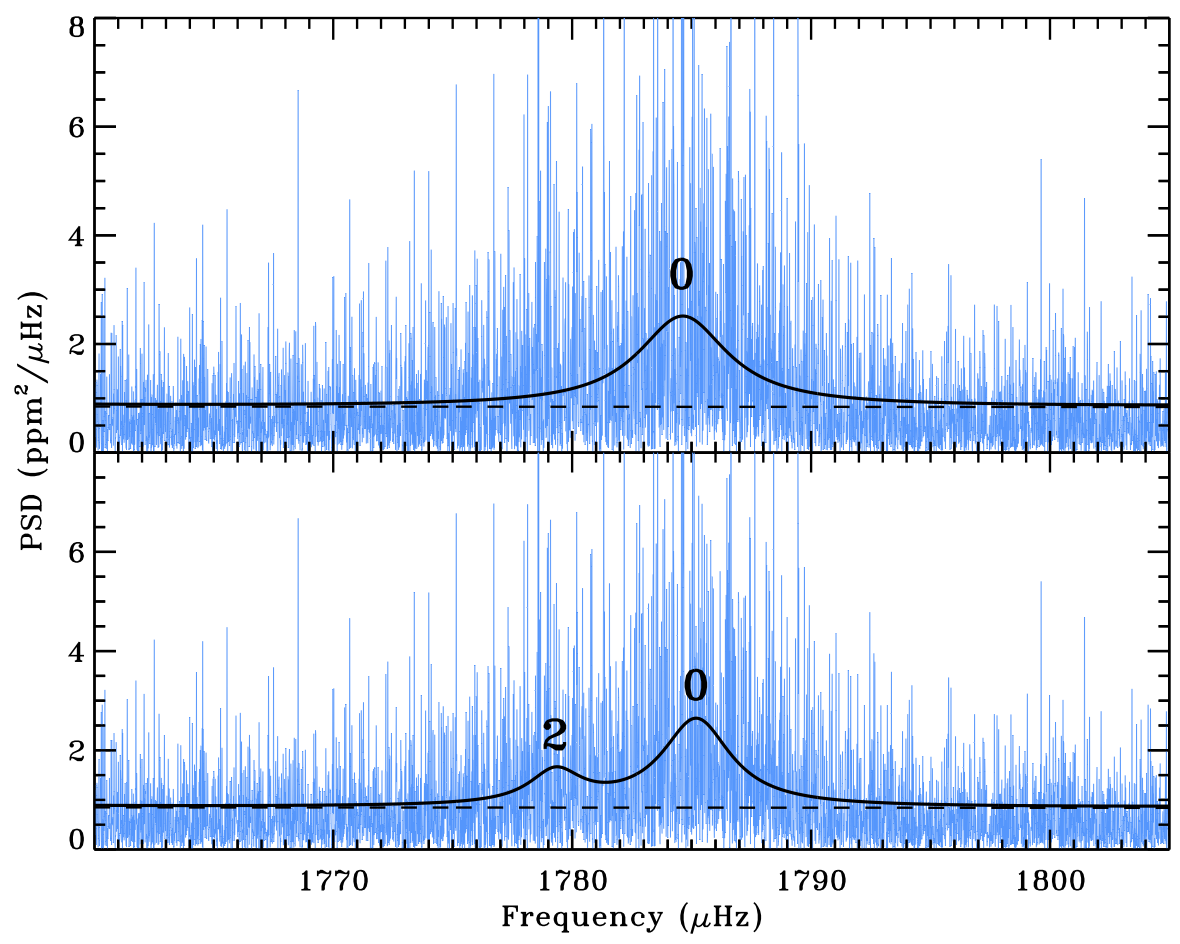}}
\caption{Background fitting for two different models (left) indicated by the green line and an example of peak detection process (right) for the star KIC 9139163. Model comparison favors the models used in the two bottom panels.}
\label{fig:bkg}       
\end{figure}

Determining how many different background signals are observed in the
stars' power spectrum (Fig.~\ref{fig:bkg}, left) can be done by means of a
model comparison based on the Bayesian Evidence, where each competing
model includes a different representation of the background level. With the
same method, one can also test the significance of an oscillation peak, in
which the competing models will either include or not the peak to be
assessed (Fig.~\ref{fig:bkg}, right). Model comparison becomes this way a very
straightforward task \cite{Corsaro14}.

\section{The novel approach: Multi-modality}
\label{sec:2}
\begin{figure}
\centering
\resizebox{0.43\columnwidth}{!}{%
 \includegraphics{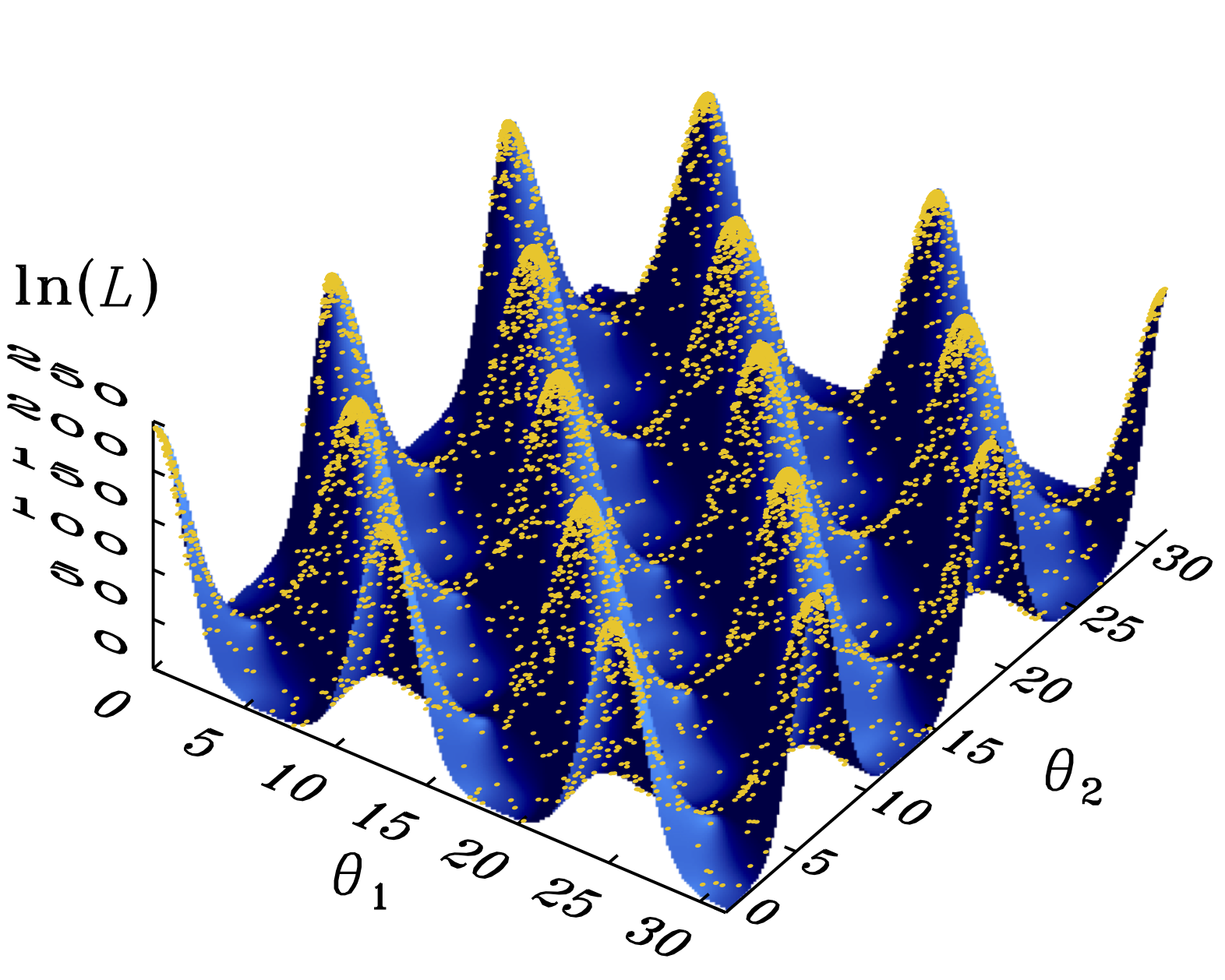} }
\caption{Example of multi-modal distribution, the Eggbox function. Samples are shown by yellow dots.}
\label{fig:eggbox}       
\end{figure}

Conversely to other existing sampling methods (e.g. based on Markov Chain algorithm \cite{Handberg11}) \diamonds\,\,allows to sample highly multi-modal distributions very efficiently \cite{Corsaro14}. The Eggbox function (Fig.~\ref{fig:eggbox}) is an example of degenerate (multiple) solution, namely a posterior probability distribution with several modes (hence multi-modal), sampled by \diamonds. We exploited the multi-modality as a novel approach to the peak bagging, succeeding in reducing the number of free parameters used to fit 27 consecutive oscillation peaks from 81 (a Lorentzian profile for each peak, hence 3 free parameters) to only 9 free parameters in total. The approach is very fast and efficient and is very well suited for automatizing the peak bagging analysis for future applications to several oscillating main sequence stars.

\end{document}